\documentclass[conference]{IEEEtran}
\IEEEoverridecommandlockouts
\usepackage{cite}
\usepackage{url}
\usepackage{amsmath,amssymb,amsfonts}
\usepackage{algorithmic}
\usepackage{graphicx}
\usepackage{textcomp}
\usepackage{xcolor}
\def\BibTeX{{\rm B\kern-.05em{\sc i\kern-.025em b}\kern-.08em
    T\kern-.1667em\lower.7ex\hbox{E}\kern-.125emX}}

\usepackage{multirow}
\usepackage{adjustbox}
\usepackage{array}
\usepackage{diagbox}
\usepackage{url}

\usepackage{tabularx}

\makeatletter
\newcommand{\linebreakand}{%
  \end{@IEEEauthorhalign}
  \hfill\mbox{}\par
  \mbox{}\hfill\begin{@IEEEauthorhalign}
}
\makeatother

\begin{document}

\title{Protection of Vulnerable Road Users using \\ Hybrid Vehicular Networks}

\author{\IEEEauthorblockN{Oscar Amador Molina}
\IEEEauthorblockA{
\textit{Halmstad University}\\
Halmstad, Sweden \\
oscar.molina@hh.se}
\and
\IEEEauthorblockN{Erik Ronelöv}
\IEEEauthorblockA{\textit{AstaZero AB}\\
Gothenburg, Sweden \\
erik.ronelov@astazero.com}
\and
\IEEEauthorblockN{Katarina Boustedt}
\IEEEauthorblockA{\textit{AstaZero AB}\\
Gothenburg, Sweden \\
katarina.boustedt@astazero.com}
\linebreakand 
\IEEEauthorblockN{Jesper Blidkvist}
\IEEEauthorblockA{\textit{AstaZero AB}\\
Gothenburg, Sweden \\
jesper.blidkvist@astazero.com}
\and
\IEEEauthorblockN{Alexey Vinel}
\IEEEauthorblockA{
\textit{Karlsruhe Institute of Technology}\\
Karlsruhe, Germany \\
alexey.vinel@kit.edu}
}
\IEEEoverridecommandlockouts
\IEEEpubid{\makebox[\columnwidth]{978-1-6654-7698-0/22/\$31.00~\copyright2022 IEEE \hfill} \hspace{\columnsep}\makebox[\columnwidth]{ }}

\maketitle

\IEEEpubidadjcol

\begin{abstract}
The use of reactive detection technologies such as passive and active sensors for avoiding car accidents involving pedestrians and other Vulnerable Road Users (VRU) is one of the cornerstones of Cooperative, Connected, and Automated Mobility (CCAM). However, CCAM systems are not yet present in all roads at all times. The use of currently available technologies that are embedded in smartphones, such as location services and Internet access, are enablers for the early detection of VRUs. This paper presents the proof-of-concept of a system that provides vehicles with enough information about the presence of VRUs by using public cellular networks, an MQTT broker, and IEEE 802.11p-enabled hardware (a roadside unit and an on-board unit). The system was tested in an urban environment and in a test track, where its feasibility was evaluated. Results were satisfactory, proving the system is reliable enough to alert of the sudden appearance of a VRU in time for the driver to react.
\end{abstract}


\section{Introduction}
According to the World Health Organization (WHO), 1,282,150 people died in 2019 in road accidents around the world~\cite{Who2021}. From those fatalities, more than half are in the Vulnerable Road User (VRU) category~\cite{Who2022}. In Europe alone, that same year, 22,756 people died in traffic accidents, and pedestrian and cyclists (i.e., VRUs) made up 20.2\% and 8\% of those fatalities respectively. This has triggered efforts such as Vision Zero~\cite{ecVisionZero} by the European Commission (EC), which includes policies stemming from law enforcement and education to the use of technology to reach the goal of having zero deaths and zero serious injuries in traffic accidents by the year 2050.

In terms of technology, the use of Intelligent Transport Systems (ITS) to enable Cooperative, Connected, and Automated Mobility (CCAM) is one of the pillars being built by the European Union to support Vision Zero. The European Telecommunication Standards Institute (ETSI) has developed a set of ITS standards that form the ETSI ITS protocol stack, which currently supports different access layer technologies but has a mature specification for ITS-G5 --- a dedicated short-range communication (DSRC) based on WiFi (802.11p). There are also efforts from  industry and academia to use cellular technologies for vehicular safety applications~\cite{Soto2022} --- C-V2X, where C stands for cellular and V2X represents the vehicle-to-anything communication capabilities. Vision Zero~\cite{ecVisionZero} considers the use of both WiFi and cellular based technologies on early stages, and acknowledges emerging technologies such as 6G for later stages~\cite{Mizmizi2021}.

However, the Vision Zero scenario where CCAM is available in all roads and at all times is still on the horizon. Even with widely available and accepted technologies such as cellular, the deployment of 5G might take some time before it becomes the dominant technology, not only in terms of network availability but more importantly in user equipment (UE) pervasiveness. The International Telecommunication Union (ITU) 2021 Information and Communication Technologies: facts and figures report~\cite{ictfactsfigs}, shows it took six years (from 2015 to 2021) for LTE surpass 3G as the dominant technology in 2021, after being launched in 2009. With the first publicly available 5G network having been launched in 2019, there is still a long road before most of the existing UEs are 5G-enabled, where V2X (or P2X, pedestrian-to-anything) may be among the most challenging capabilities.

Nevertheless, solutions involving the use of available technologies such as currently deployed mobile-cellular networks in combination with existing ITS infrastructure could provide immediate benefits. In this paper, we propose a system that uses mobile cellular networks to provide safety to VRUs by making them visible to drivers even without a physical line of sight. In our system, a VRU uses a mobile app to publish its position and time information (POTI) in an MQTT broker. From there, a middle-ware reads the POTI information and prompts a Road Side Unit's (RSU) API to send a Personal Safety Message (PSM) as defined by Society of Automotive Engineers (SAE) J2735\_202007~\cite{SAEdictionary}. The message is then broadcast to ITS-enabled vehicles nearby that react to the presence of the VRU. This setup is the first stage in the development of a standalone VRU protection device that is compliant with either the ETSI or the SAE specifications. At this point (proof-of-concept --- POC), results show that there are still stages to go through before the standalone device reaches the minimum viable product (MVP) stage.

The contributions of this paper are:
\begin{itemize}
    \item The presentation of a POC for a hybrid system (LTE and DSRC) for VRU protection.
    \item Test of the POC system in an urban scenario and in a test track
    \item A discussion of the future stages for the system considering design  principles and the availability of access technologies
\end{itemize}

The rest of the paper is organized as follows: in section \ref{sec:related_work}, we explore the current efforts in the protection of VRUs, including their inclusion in standards such as ETSI ITS and SAE. In Section \ref{sec:system_description}, we describe our current deployment and the projected final version of the standalone VRU protection device. In Section \ref{sec:demo}, we describe the experiment where we performed a proof-of-concept for the VRU app. Subsequently, in Section \ref{sec:discussion}, we analyze the outcome of the proof-of-concept and the future path for the project. Finally, we present our conclusion in Section \ref{sec:conclusion}.

\section{Related Work}
\label{sec:related_work}
The protection of VRUs has been present in the field of VANETs since its inception. One of the main concerns has been the role of VRUs in the network, since they can be detected \textit{passively} or \textit{actively}. An example of passive detection is the use of cameras, radar, or LiDAR to detect VRUs and other entities in the road. On the other hand, active detection occurs when the VRU makes other road users aware of its presence by using a network-enabled beacon (i.e., not simply a light or a reflective vest). Since our POC falls into the active category, we will explore works related to active detection in this section.   

One early example of active detection appears in \cite{LillKoTag}. In this work, the authors propose the use of tags carried by VRUs that have the ability to connect to other VRUs and vehicles' OBUs. They consider the dynamics of VRUs and the fact that tags must be energy efficient.The authors postulate it as an extension to the 802.11p protocol, and reach the simulation stage for their proposal. As opposed to our current POC, they do not consider the use of mobile phones, but they explore the use of additional hardware which we are currently probing in a form different to a tag.

The work in \cite{NGUYENMEC}, however, does consider the use of mobile phones, which have the advantage of being carried by VRUs. In this work, authors propose the use of LTE-enabled phones to obtain a pedestrian's POTI information and use Multi-access Edge Computing (MEC) to calculate and inform VRUs and vehicles about the possibility of a collision. While this work presents an extensive study on the computational cost of collision detection algorithms and energy consumption in the UE, it fails to address how it intends to overcome the increasing hurdles set by mobile operating systems --- namely Android and iOS --- for developers to access precise POTI information at all times due to users' concerns about privacy. We explore this limitation in Section~\ref{subsec:system_limitations}.

Furthermore, the same team explores the requirements for their collision algorithm in \cite{NguyenReqs}. In this work, the authors consider the use of activity detection (i.e., sensing whether the user is walking, standing, or crossing a curb) to aid its collision detection algorithm. They conclude that for POTI inaccuracies between 0.5\,m and 1.0\,m, delays must not exceed 100\,ms and 300\,ms for activity detection and communication in order for their algorithm to be effective. Nevertheless, they reach only the simulation stage, and they fail to consider the limitations of location services in smartphones, which we address in Section \ref{subsec:demo_urban}. 

\section{System Description}
\label{sec:system_description}

\subsection{System Architecture}

\begin{figure}[tb!]
	\centering
	\includegraphics[width=\textwidth/2]{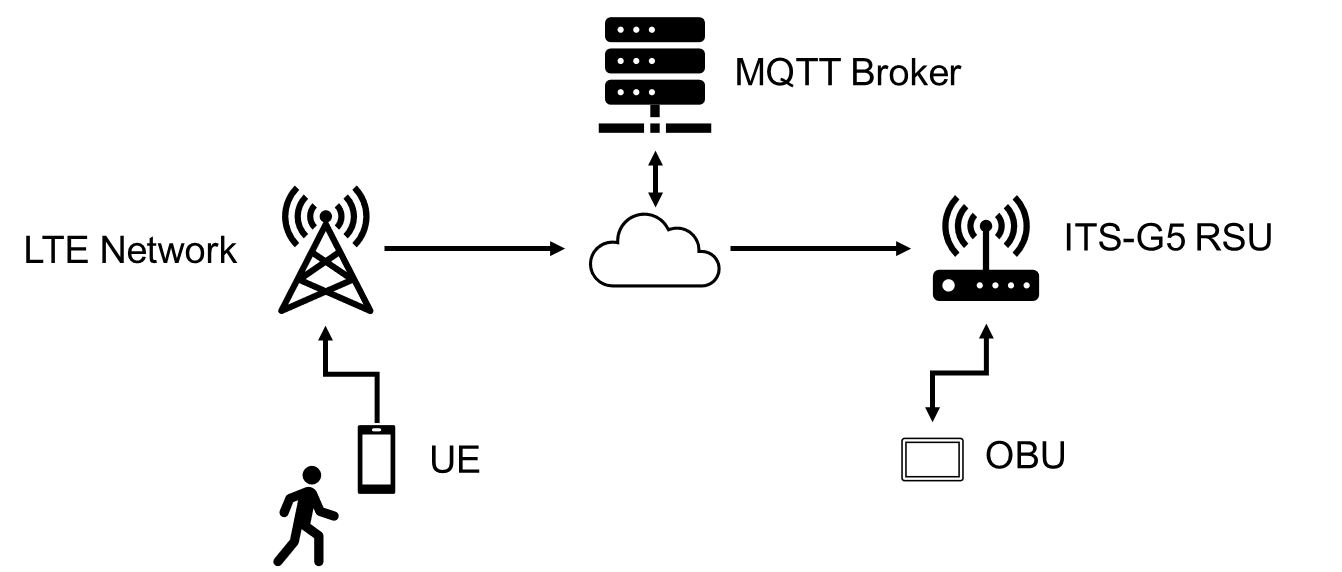}
\caption{System Diagram}
\label{fig:system_arch}
\end{figure}

The initial use case for the system falls within the scope of AstaZero~\cite{AstaZero} --- a full-scale automotive test environment based in Sweden. Because of its independent nature, many companies perform tests at the \mbox{AstaZero} tracks. This means that, at any given time, test engineers and other track users are present on the test roads and hence in a vulnerable position. Currently, \mbox{AstaZero} provides a safety positioning service for vehicles which uses an MQTT broker, which can be extended by the proposed system.

Fig.~\ref{fig:system_arch} shows the main actors in the system: the VRU, the infrastructure (i.e., MQTT broker, interface, and RSU), and the ITS-equipped vehicle. For this POC, a mobile application runs on the VRU mobile phone equipped with Android or iOS. Using the native location services, the application publishes its POTI information onto the \mbox{AstaZero} MQTT broker. POTI information is not published periodically, since it follows a logic similar to that of the ETSI Cooperative Awareness Message (CAM)~\cite{etsiCAM}, where messages are triggered according to the ITS station (ITS-S) dynamics, such as significant changes in position~\cite{Lyamin2017}. 

The information in the broker is consumed by an \mbox{AstaZero} proprietary middle-ware which connects to a commercially available RSU through its API. Using the POTI information from the MQTT message, the RSU crafts a message. For the purpose of this POC, the message is a SAE PSM~\cite{SAEdictionary}, but the RSU can generate ETSI messages as well, such as Decentralized Environmental Notification Messages (DENM)~\cite{etsiDENM}. A thorough discussion on which message provides more advantages is provided in Section~\ref{sec:discussion}. 

The message is broadcast from the RSU and received by its neighbors at a one-hop distance. In this POC, a vehicle is equipped with a commercially available on-board unit (OBU) which receives and processes the message and displays the received POTI information as an icon in a tablet screen. This alert in the human-machine interface (HMI) prompts the driver to react accordingly to the presence of a VRU.

\subsection{System Requirements}
System requirements can be divided in two categories: i) technical performance requirements, and ii) product-related requirements. For the case of i), the system must provide accurate information in a reliable fashion. For ii), \mbox{AstaZero} provided a set of features that the mobile application must offer.

The accuracy of the proposed system consists of providing the location of an agent within a certain margin. Location services that are built into today’s smartphones are marketed as accurate within meters when the global navigation satellite system (GNSS) antenna inside the mobile phone is used. The work presented in~\cite{CastroGPS} states the accuracy of location information from smartphones is within the 3 to 6 meters boundary, whereas authors in~\cite{LeeGPS} measure the performance of mobile location services in phones and tablets from different manufacturers and obtain precision values ranging from 0.465\,m to 72.28\,m.

Reliability refers to the ability of obtaining the information from the server as soon as possible. This means that the vehicle will be able to detect target VRU with enough time to react. The document in~\cite{UKcode} provides typical stopping distances in a regular situation regarding road surface, weather, and mechanical condition of the car. It considers two different times, a typical \textit{Thinking Distance} which is the time between noticing the risk and starting to brake, and a \textit{Braking Distance}, which is the time it takes for the braking maneuver to complete. An example from~\cite{UKcode} demonstrates that for a typical car traveling at 64\, km/h in a straight road, it would take 3.18 seconds for a car to reach a full stop from the moment the driver became aware of the danger, with 0.66 seconds being the thinking time and 2.52 seconds being the time it takes the car to brake. This would occur in a distance of up to 36\,m before reaching a full stop. Of course, these figures change when we consider automated driving and when advanced driver-assistance systems (ADAS) are present. However, this means that the system has to be reliable enough to deliver at least one accurate message 3.18\,s before the estimated time-to-collision (TTC).

Regarding product requirements, \mbox{AstaZero} provided a set of features that the mobile app must cover, namely:
\begin{itemize}
    \item Geofencing: the app must only operate within the \mbox{AstaZero} proving grounds.
    \item Determine if a user is a VRU: the app must pause publishing POTI information if the user stops being vulnerable (e.g., when the user is moving inside a vehicle).
    %\item Two-way alerts: this optional feature would use the haptic feedback in the UE to alert the VRU about approaching vehicles.
\end{itemize}

Both Android and iOS provide libraries that enable the requirements set by \mbox{AstaZero}. The POC currently complies with the mandatory features: it has a circular geofence centered in the \mbox{AstaZero} grounds, and it uses the activity sensors (e.g., accelerometer and gyroscope) to help determine the degree of vulnerability of a user.

\subsection{System Limitations}
\label{subsec:system_limitations}
At the POC stage, parts of the system exhibit some limitations, specifically the mobile app and the MQTT broker. The mobile app performance is limited by the access to resources present in the UE, namely location services. Due to the behavior of the Android and iOS operating systems, the location services are not able to provide the most updated POTI information unless the app is running in the foreground (i.e., the screen cannot be locked). Furthermore, the MQTT broker can potentially become a single point of failure, which in this case can be solved by adding redundancy. These issues are taken into account for future stages of this project, after the MVP stage, where the intention is to have a standalone device within a decentralized network, i.e., a pure vehicular ad hoc network (VANET).

\section{Demonstration}
\label{sec:demo}
We performed a demonstration for the POC in an urban environment and at the \mbox{AstaZero} proving ground. The test in an urban environment was performed around the \mbox{AstaZero} headquarters, at Lindholmen Science Park in the Gothenburg metropolitan area, Sweden. The \mbox{AstaZero} proving grounds are located in Sandhult, Sweden. 

The mobile application was installed in an Android and in a iOS device. The Android application runs on a Google Pixel 6 phone, and the iOS app runs on a second-generation iPhone SE. Both phones used a Telenor SIM card for each trial run. Finally, both UEs used the fused location services available for Android and iOS, which means that they request the most precise location available and the operating system uses GNSS, the mobile network, and other services to retrieve the most precise positioning data. 

The V2X hardware consisted of commercially available devices: an RSU and an OBU. The RSU was connected to the \mbox{AstaZero} network and located in a window facing the street in the urban experiment, and in the edge of the test area for the trial in the proving grounds. The OBU was mounted in the back of a 2021 Lynk \& Co 01 SUV, with a tablet set in the cockpit in the place of the rear view mirror to serve as the HMI. 

\subsection{Urban Environment}
\label{subsec:demo_urban}

\begin{figure}[tb!]
	\centering
	\includegraphics[width=\textwidth/2]{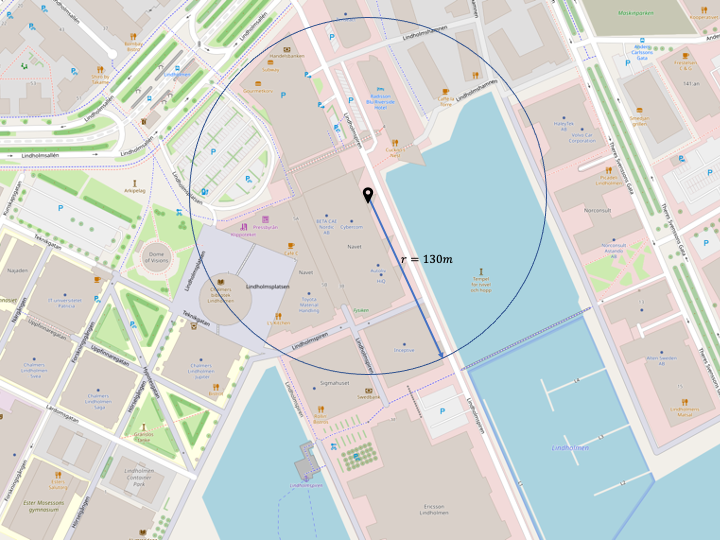}
\caption{Urban Environment: RSU (black pointer) and the coverage radius over Lindholmspiren street}
\label{fig:urban_map_coverage}
\end{figure}

For the urban environment, we performed a test to try the RSU coverage and the overall functioning of the system. We drove the vehicle around the area, specifically along the streets Lindholmspiren, Lindholmsallén, and Theres Svenssons gata. These areas contain medium to high rise buildings, some of them housing research facilities, factors which we expected to affect the DSRC connection. Fig.~\ref{fig:urban_map_coverage} shows the location of the RSU and the zones along Lindholmspiren where the RSU and OBU started exchanging information, which is a distance of approximately 130\,m to either side of the RSU along Lindholmspiren. The distance between the RSU and Theres Svenssons gata  exceeds 130\,m, so the OBU did not receive information from the RSU at that distance.

\begin{figure}[tb!]
	\centering
	\includegraphics[width=\textwidth/2]{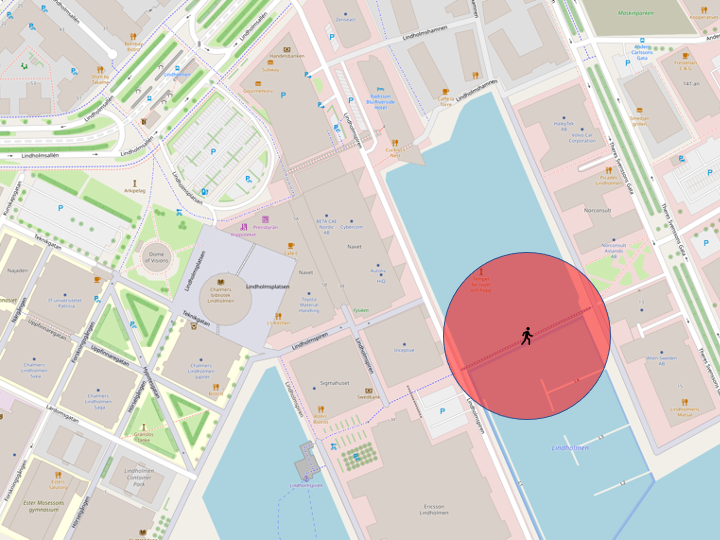}
\caption{Urban environment: the actual VRU location (black icon) and the area within which location information fluctuated (red)}
\label{fig:urban_gps_variation}
\end{figure}

The mobile application maintained the connection to the MQTT broker as long as the app was in the foreground. However, we noticed an effect of relying on the phone location provider: when GNSS information was not available, the mobile phone would send an estimated location to the MQTT broker. We explored this phenomenon by walking to the middle of the bridge Lindholmsbron in the middle of the water, and the HMI in the vehicle would show the pedestrian on the far side of the bridge. We attribute this to the fact that the location provider switched from GNSS to positioning provided by the mobile network. This prompted us to mitigate this problem before the test at the proving grounds by using a rolling average of the last positions, and to consider a different solution (e.g., discard position changes exceeding a threshold) for future releases. These inaccuracies in location are consistent with the results in \cite{CastroGPS} and \cite{LeeGPS}. Fig.~\ref{fig:urban_gps_variation} shows an example of these variations, with the actual and the provided positions.

\subsection{Test Track}
\label{subsec:demo_track}

After correcting the issues on the mobile app, we evaluated the system in the \mbox{AstaZero} proving grounds. In a rectangular field, we had a set of vehicles, two sedans and a bobtail truck. The RSU is outside the rectangular field but within the range for communication with the RSU. A pedestrian is walking behind the truck, equipped with the mobile app that is publishing its location to the MQTT broker. The trial consists on confirming that the OBU receives the PSM alerting of the presence of a VRU before the pedestrian is seen by the driver. A video showing the system in operation is available at \url{https://youtu.be/fcGm35zgVU8}

\begin{figure}[tb!]
	\centering
	\includegraphics[width=\textwidth/2]{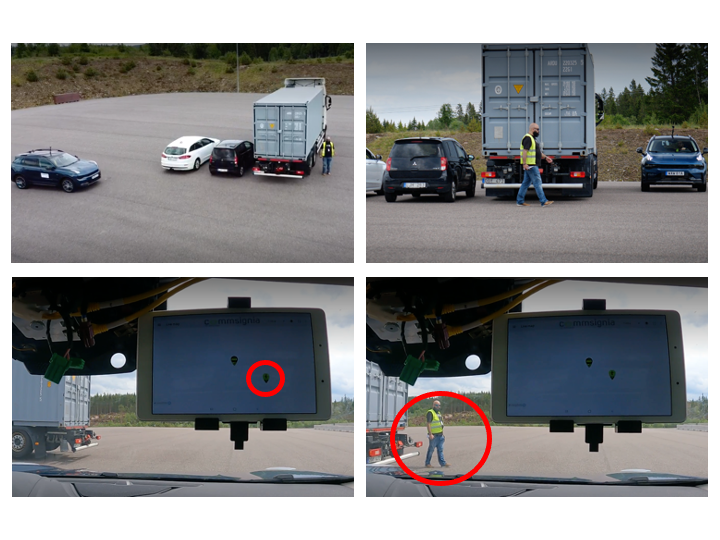}
\caption{Test Track: above, testing scenarios; and below two frames showing the HMI (left) alerting about the VRU before it is visible (right)}
\label{fig:photo_track}
\end{figure}

During this test, the location provided to the MQTT broker exhibited less variations. Furthermore, since this specific test track is out in the open, GNSS is naturally more precise. Fig.~\ref{fig:photo_track} shows the scenarios we tested and a frame showing that the HMI displays the VRU location before it is in line of sight.

Considering both experiments, we can conclude that a system that uses a device connected to an MQTT broker as a bridge to a VANET is feasible, and that it has the potential to provide a driver or an automated vehicle with the necessary information to avoid a collision. In Section~\ref{sec:discussion}, however, we will discuss the advantages and disadvantages of relying on commercial off-the-shelf UEs as opposed to standalone devices, as well as the potential pros and cons of this hybrid system when compared with standardized VANETs. 

\section{Discussion}
\label{sec:discussion}

The validation of the proposed system poses a set of questions and challenges for the future steps, namely: issues stemming from the use of commercial off-the-shelf UEs, potential single points of failure and scalability, and compliance with standards. In this section, we discuss the future steps for the project, including its possibility to reach widespread implementation.

\subsection{On the use of commercial mobile phones}
The use of commercial off-the-shelf UEs to develop and test this POC offers the possibility for distributing the application through several channels, specially the official Android and iOS application marketplaces. For our initial use case, that would mean that the users at the \mbox{AstaZero} proving grounds would only need to download the application once when checking in. This would allow the company to enable VRU safety without the need to purchase UEs or another hardware. Nonetheless, this would require for security measures to be adopted, e.g., creating a single-use token to restrict the use of the app only to track users that have checked in for the day.  

However, the use of mobile phones adds to the limitations for the system, as explored in Section~\ref{subsec:system_limitations}. As of today, the access to location services is restricted by the mobile operating systems, and precise locations can only be accessed if: 1) the screen is unlocked, 2) the application is running in the foreground, and 3) the user has explicitly granted the app permission to access the location services. This last point is crucial, since end users can revoke permissions at any time whether intentionally or accidentally, rendering the application useless. 

Considering the criticality of the system, and in order to expedite its deployment in the \mbox{AstaZero} grounds, our next step towards reaching the MVP stage is to use an LTE-enabled single-board computer. These computers would be available for users at check-in, most likely mounted on the safety vests that are provided before entering the grounds. This would ensure that the system performance is under company control, and would not depend on the varying characteristics of the users' devices (e.g., while iOS devices have very similar characteristics, Android devices have a wide range of hardware features). Finally, the use of single-board computers enables the ability to pivot towards other versions of the system, e.g., change the access technology from LTE to DSRC technologies such as ETSI ITS-G5 or WAVE.

Among the challenges brought by the use of single-board computers, power consumption is the one we detect a priori. We are currently testing a SolidRun HummingBoard  computer, which operates at 7-36\,V, out of range for USB 5\,V power banks. Besides, the use of additional hardware (e.g., GNSS, gyroscopes, accelerometers, DSRC network interface cards) would put more pressure on battery consumption. These issues are to be addressed during the MVP stage, where a usable system is to be proposed.

\subsection{On single points of failure and scalability}
For the initial use case, the \mbox{AstaZero} MQTT broker is a potential single point of failure. The only connection between the VRU and the VANET (RSU and OBU) goes through the broker and the interface. This can be solved by adding redundancy. Also, the MQTT broker is a crucial part of the existing safety system in the grounds, so if the broker fails, safety protocols are in place to ensure the safety of \mbox{AstaZero} users.

Scalability is a potential issue even within the \mbox{AstaZero} scope. The proving ground contains five full-size test tracks: a multi-lane area, a rural road, a city area, a high-speed zone and an indoor track (dry zone). This means that the MQTT broker should interface with several RSUs and either the interface or the RSU should decide which messages are relevant and only broadcast those (e.g., a VRU in the city area should not be broadcast to the multi-lane area). This would add to the processing time, and affect the end-to-end delay between message triggering and reception, which is crucial if the RSUs broadcast messages just to one-hop neighbors.

Both issues can be tackled by using pure VANETs, e.g., running ETSI ITS protocol stacks on a single-board computer. This way, the VRU can broadcast its location directly to the VANET side of the system, either to the RSU or directly to the OBU. This would also allow for the system to be used in other deployments compliant with ETSI ITS.

\subsection{On the compliance with ITS standards}

The POC as presented is not fully compliant with standards. The SAE specification~\cite{SAEpsm} states that PSMs shall be generated by handheld devices carried by  pedestrians, bicycle riders and public safety personnel. However, in this POC, it is the RSU broadcasting the PSMs, as it was the case in the system proposed in~\cite{IslamPSM}. This logic is also followed by the ETSI-defined Vulnerable Road User Awareness Message (VAM)~\cite{etsiVAM}, where it is the VRU or the cluster head of a group of VRUs that transmit a message, not an RSU. 

To make the current system compliant, for example, with the ETSI ITS standard, an adjustment can be made so that the RSU transmits a DENM. One of the cause codes for DENMs is human presence on the road~\cite{Campolo2019}. There is also an advantage of DENMs, which is that other stations can forward the message in order to cover a destination area, as opposed to single-hop broadcasting where the coverage area is limited by the one-hop transmission range of a ITS station. We have explored the use of multi-hop dissemination of safety messages in \cite{Amador2022}, where we evaluate the performance of the ETSI GeoNetworking specification~\cite{etsiNewGeoNetworking} to distribute DENMs in a geographical area. Additionally, using DSRC-enabled single-board computers, the VRU itself would be able to send these DENMs directly to the vehicles and use them, alongside the RSUs, to disseminate them at multiple-hops, making stations several kilometers away aware of the presence of VRUs on the tracks, mitigating the risk of not detecting a VRU on time.

Just as importantly, full compliance with standards ensure that the system can escalate and be deployed, with little to no effort, in other compliant setups. E.g., if the DSRC-enabled single-board computer is fully compliant with ETSI ITS, it can be used in any ETSI-compliant deployment anywhere else in Europe, not only within the \mbox{AstaZero} grounds. This would mean that the final product, or even an advanced MVP, would improve the safety of VRUs in a relatively short time.

\section{Conclusion}
\label{sec:conclusion}
We presented a POC for a system relying on hybrid network technologies (cellular and DSRC) to enable VRU protection. The demonstration showed that mobile phones applications would work better in closer environments where system owners can alter the operating system to tap into the full capabilities of smartphones. However, larger deployments require the use of standard-compliant hardware that is able to operate in full capacity and at all times. While the POC was successful, the discussion renders an embedded system as the go-to for the MVP stage.

\section*{Acknowledgment}

Halmstad University acknowledges the support from Knowledge Foundation ("Safety of Connected Intelligent Vehicles in Smart Cities -- SafeSmart" project), from VINNOVA ("Emergency Vehicle Traffic Light Preemption in Cities -- EPIC" project 2020-02945) and from ELLIIT Strategic Research Network. 

AstaZero acknowledges the support from VINNOVA ("V2X and connected infrastructure -- V2X2" project 2019-05899). 

Karlsruhe Institute of Technology acknowledges the support from Helmholtz Association (FE.5167.0007.0012 -- Cooperative Autonomous Systems).

\bibliographystyle{IEEEtran}
\bibliography{mybibfile}

\end{document}